\documentclass[pdftex,twocolumn,epjc3]{svjour3}          

\RequirePackage[T1]{fontenc}

\smartqed  

\usepackage[short]{optidef}
\usepackage{algorithm}
\usepackage[noend]{algpseudocode}
\usepackage{amsmath}
\usepackage{amssymb}
\usepackage{float}
\usepackage{color} 
 \usepackage{fancyvrb}

\RequirePackage{graphicx}
\RequirePackage{mathptmx}      
\RequirePackage{flushend}
\RequirePackage[numbers,sort&compress]{natbib}
\RequirePackage[colorlinks,citecolor=blue,urlcolor=blue,linkcolor=blue]{hyperref}

\journalname{International Journal of Information Security}

\begin{document}

\title{Cyber-Physical Risk Assessment for False Data Injection Attacks Considering Moving Target Defences \thanksref{t1}}

\subtitle{Best practice application of respective cyber and physical reinforcement assets to defend against FDI attacks.}

\author{Martin Higgins \thanksref{addr1,addr2,e1}
        \and
        Wangkun Xu \thanksref{addr1,e2},
         Fei Teng \thanksref{addr1,e3},
         and 
        Thomas Parisini \thanksref{addr1,e4}
}

\thankstext[$\star$]{t1}{This work was partly supported by ESRC under Grant ES/T000112/1 and EPSRC Centre for Doctoral Training in Future Power Networks and Smart Grids (EP/L015471/1). Also, this work has been partially supported by the European Union's Horizon 2020 research and innovation programme under grant agreement No 739551 (KIOS CoE) and by the Italian Ministry for Research in the framework of the 2017 Program for Research Projects of National Interest (PRIN), Grant no. 2017YKXYXJ. (\textit{Corresponding author: Fei Teng})} 
\thankstext{e1}{martin.higgins@eng.ox.ac.uk}
\thankstext{e2}{wangkun.xu@ic.ac.uk}
\thankstext{e3}{f.teng@ic.ac.uk}
\thankstext{e4}{t.parisini@ic.ac.uk}

\institute{Imperial College London, South Kensington, London, UK\label{addr1}
          \and
          University of Oxford, Osney Mead, Oxford, UK\label{addr2}
}

\date{Received: date / Accepted: date}

\maketitle

\begin{abstract}
In this paper, we examine the factors that influence the success of false data injection (FDI) attacks in the context of both cyber and physical styles of reinforcement. Exsiting research considers the FDI attack in the context of the ability to change a measurement in a static system only. However, successful attacks will require first intrusion into a system followed by construction of an attack vector that can bypass bad data detection (BDD) to cause a consequence (such as overloading).Furthermore, the recent development of Moving Target Defences (MTD) introduces dynamically changing system topology, which is beyond the capability of existing research to assess. In this way, we develop a full service framework for FDI risk assessment. The framework considers both the costs of system intrusion via a weighted graph assessment in combination with a physical, line overload-based vulnerability assessment under the existence of MTD. We present our simulations on a IEEE 14-bus system with an overlain RTU network to model the true risk of intrusion. The cyber model considers multiple methods of entry for the FDI attack including meter intrusion, RTU intrusion and combined style attacks. Post-intrusion, our physical reinforcement model analyses the required level of topology divergence to protect against a branch overload from an optimised attack vector. The combined cyber and physical index is used to represent the system vulnerability against FDIA.

\end{abstract}

\section{Introduction}
The contemporary power system is a cyber-physical system with high levels of system inter-dependency and a near ubiquitous use of communications throughout. The move towards cyber-physical systems has resulted in new vulnerabilities which have not been fully covered by the existing defense frameworks. Such vulnerabilities were exposed during the 2015 cyber-attack against distribution companies in Ukraine \cite{Liang2017TheAttacks}. Attacks like these have increased the focus on the area of power system cyber security. While many papers have focused on designing new attacks and novel defences, relatively few have focused on risk assessment of specific cyber attack types within the context of a cyber-physical system. The main contribution of this work is to provide a cyber-physical model of risk assessment for FDI attacks. This model considers the inherent risk of a system topology, the interconnection between RTU and telemetered measurements. The model also considers post-intrusion attack considerations of attack plausibility in light of the active defence technique, Moving Target Defence (MTD).

\section{Related Works}

\subsection{FDI Attacks}

False Data Injection (FDI) attacks, were first outlined in \cite{Liu2011FalseGrids} and involve altering system measurements to corrupt a network operator's state estimation process and cause negative consequences such as line overloading and outage masking \cite {Liu2016MaskingAttacks}. FDI attacks need to remain undetected by the network operator to be effective. In this context, FDI attacks compete with bad data detectors (BDD) within state estimation processes. In modern energy management systems (EMS), the BDD at the power system level relies on weighted-least squares (WLS) and chi-squared error testing \cite{Liang2017ASystems}. Therefore, a successfully FDI attack can only be implemented if the attacker gains access to the correct combination of meter measurement and change them  in a coordinated way. Deng \textit{et al}. offer a  comprehensive review of the FDI attack problem in \cite{Deng2017FalseSurvey}.

\subsection{Moving Target Defence}
Moving Target Defence (MTD), within the realm of power system FDI attacks, refers to the process of imposing dynamic changes in the physical system to invalidate the topology knowledge assumption of the FDI attacker. As we have shown previously in \cite{Higgins2021TopologyInformation}, it is possible to extract network topology and perform FDI attacks simply by observing power system data. By introducing topology changes away from expected model by the attackers, gross errors can be introduced which expose attackers via residual violations. These topology changes can be implemented through transmission switching \cite{Wang2015EffectsNetworks} (rarely suggested) or by admittance perturbation via distributed flexible AC transmissions (D-FACTs) devices \cite{Morrow2012TopologyInjection}. Extensive research has been carried out for the optimisation of MTD application for power systems \cite{Liu2017ReactanceEstimation} \cite{Li2019OnDevices}. Other areas of research have been on advancing MTD by camouflaged or hidden MTD \cite{Tian2019EnhancedGrids}  \cite{Higgins2020StealthySystems} or by exploring the cost of applying MTD \cite{Lakshminarayana2020Cost-BenefitGrids}. In response to this cost-based analysis, other works have also considered the implementation of MTD via event-triggering to reduce the overall utilisation of MTD \cite{Higgins2020EnhancedDefence}. Crucially, no previous research have yet considered the application of this style of MTD from a risk assessment perspective. Questions remain on when to apply MTD and when to opt for traditional cyber style reinforcement. In this work we would like to address, how resources can be best spent in cyber-physical networks to reduce the overall risk of a successful FDI attack (in both entry and post-intrusion detection terms). While many works have discussed the completeness of MTD, the ability to protect a network with MTD under noisy environment will be dependent on the size of the attack vector. If a network is regularly operating at or close to its rated capacity, then the attacker can cause damage with only minor changes to the system. Under this scenario, large amounts of MTD will be required in order to drive a positive detection, which may be potentially untenable and may make cyber reinforcement the more effective solution.

\subsection{Cyber-Physical Risk Assessment}
Some works have already attempted to tackle risk assessment with respect to FDI attacks. For example, Hug \textit{et al}. examined this area from the perspective of weakest node attack point \cite{Hug2012VulnerabilityCyber-attacks}. In this work they perform a nodal based target selection using a minimum meters criteria to compromise a node or state angle. The number of meters required for each node are evaluated for the AC and DC models and an alternate meter conquering strategy is proposed using the RTUs rather than the individual meter measurements.
A similar methodology is explored in \cite{SandbergOnNetworks} where security indices are developed based on the physical topology of the power system with specific reference to the FDI attacks. In this case, the security index is defined by the minimum meter change potential with an aim of finding the sparsest possible attack. However, the work makes no reference to the RTU or combination style vulnerabilities. A similar indices-based approach is also applied in \cite{Teixeira2015SecureApproach}. In \cite{Pan2019CyberEstimation} Pan \textit{et al}. offer one of the first risk assessments of FDI attacks with cyber considerations. The attack combines standard FDI style attack vectors with denial of service (DoS) style attacks which reduces the number of meters required to compromise a state. 


Some other works have addressed cyber-physical risk modelling more generally. In \cite{Wu2016RiskVulnerabilities} a probabilistic risk assessment model is introduced. The model uses acyclic digraphs to represent the inter-dependencies between different components in a cyber-physical system. 
In \cite{Chopade2011CriticalNetworks} a probabilistic risk approach is used but with a focus on the removal of graph nodes or edges and the effects they have on the network. 
In \cite{Davis2015AInfrastructures} a framework for cyber-physical modelling of power grid infrastructure is outlined. The attack focus is around circuit breaker control and de-energising certain areas of the grid. They combine upper level RTU modelling with a lower level telemeter network model. One of the earliest relevant works in the field,  Bargiela \textit{et al} explore network observability as a function of network topology in \cite{Bargiela1986OBSERVABILITYTECHNIQUE}. The work also proposes an optimal protection graph which satisfies the spanning tree. This graph can then be used to return a set of optimal buses to protect and guarantee reliable state estimation. In \cite{Tantawy2020Model-BasedSecurity} an integrated model-based approach for cyber-physical risk assessment is used which outlines the vulnerabilities of specific controllers into an industrial test bed. 
In \cite{Ten2008VulnerabilitySystems} a vulnerability assessment framework for systematically evaluating SCADA vulnerabilities is proposed. The method can be used to model access points for SCADA networks, construct a model for intrusions, simulate cyber attacks and suggest security improvements. 
In \cite{Wang2016OnSecurity}, a meshed network framework that considers both power system features and bi-directional communication flows is presented. In \cite{Barrere2019IdentifyingSystems}, \textit{Barrere et al} outline a cyber-physical assessment framework which features combined style attacks for industrial control systems (ICS).

While these frameworks offer some interesting risk perspectives on the FDI attack, they can be improved in a number of ways. Consideration of overlapping style attacks, which combine RTU and meter style intrusions, would better represent the risk to a power system. Also, previous works which modelled the risk of the FDI attacks, have assumed success once the attacker has gained access to the required combination of meters. However, this fails to consider both the system state and post-intrusion defences such as MTD. It is hence important to expand upon these works by redefining a successful FDI attack with consideration of the size of required attack vector, the existence of MTD and a cyber-physical model. In the following subsection we outline the proposed risk assessment framework.

\subsection{Proposed Risk Assessment Framework}
Our cyber-physical risk assessment framework builds on the cases outlined in the literature review. In particular it considers \cite{Hug2012VulnerabilityCyber-attacks}, \cite{SandbergOnNetworks}, \cite{Pan2019CyberEstimation} \& \cite{Davis2015AInfrastructures} and attempts to create an overarching risk assessment criteria which considers both the intrusion (cyber) component of risk and stealthiness in the presence of MTD portion (physical) risks. Our work considers both the cost of intrusion to a given attack point and the ability to remain stealthy in the presence of MTD system capabilities. Table \ref{tablenovelty} outlines where our work sits within the context of the most similar papers.

\begin{table*}[]
\begin{tabular}{|l|l|l|l|l|l|l|}
\hline
Reference  & Author                 & Cyber-Physical   Model & Consideration of   FDI & Consideration of MTD & Intrusion Risk & Post-Intrusion   Risk \\ \hline
15         & \textit{Hug et al}     & X                      & \checkmark       & X             & \checkmark      & X                     \\ \hline
16         & \textit{Davis et al}   & \checkmark              & X               & X             & \checkmark      & X                     \\ \hline
18         & \textit{Pan et al}     & X                      & \checkmark       & X             & X              & X                     \\ \hline
22         & \textit{Davis et al}   & \checkmark              & X               & X             & X              & \checkmark             \\ \hline
Our work & \textit{Higgins et al} & \checkmark              & \checkmark       & \checkmark     & \checkmark      & \checkmark             \\ \hline
\end{tabular}
\caption{This reports relative novelty in the field of CPS style risk assessment for FDI style power system attacks.}
\label{tablenovelty}
\end{table*}

To this end we combine weighted min cost of intrusion modelling and the level of MTD required to protect a measurement when assessing risk to create a cyber-physical risk assessment approach. In our risk assessment framework for FDI attacks, we make the following contributions:

\begin{itemize}
    \item Firstly, our model provides a weighted graph assessment of the FDIA intrusion risk of the cyber components of the grid. We introduce overlapping style attack considerations for the FDI attack model i.e. not simply the choice of the RTU or the meter combinations for a given state but also some combinations of the two. We show in simulation, that these overlapping style vulnerabilities can reduce the attackers overall intrusion cost and hence enhance their ability to attack.  
    \item We also introduce an MTD (post intrusion) effectiveness criteria which considers system capacity constraints in the context of an FDI attack and the required level of MTD to expose an attack for an overload style attack. We model the level of divergence required to protect each bus and branch combination in the context of a min possible attack vector. We consider the MTD effectiveness in the context of statistical loading peaks and optimised attack vector.
    
\end{itemize}

The rest of this paper is organised as follows. The preliminaries are outlined in Section 3.  Section 4 details the formulation related to the cyber-physical risk model and outlines the algorithms used and their respective performances. Section 5 contains outlines for the algorithm and algorithmic performance result. Section 6 features the results and analysis of the risk model applied to different case studies and Section 7 concludes the paper.

\section{Preliminaries}
\subsection{State Estimation}

Initially, we consider the static power system problem. Consisting of a set of $n$ state variables $\textbf{x} \in \mathbb{R}^{n\times1}$ estimated by analysing a set of $m$ meter measurements $\textbf{z} \in \mathbb{R}^{m\times1}$ and corresponding error vector $\textbf{e} \in \mathbb{R}^{m\times1}$ . The non-linear vector function $\textbf{h}(\textbf{.})$ relates meter measurements $\textbf{z}$ to states $\textbf{h}(\textbf{x}) = (h_1(\textbf{x}),h_2(\textbf{x}),...,h_m(\textbf{x}))^T$ shown by 

\begin{equation}
    \textbf{z} = \textbf{h(x)} + \textbf{e}.
    \label{generalized state equation}
\end{equation}

However, we can primarily focus on the linear model for this paper as we are operating from a risk assessment perspective only. Therefore, the state estimation problem can be represented by the linear model as a function of the Jacobian $\textbf{H} \in \mathbb{R}^{m\times n}$ matrix and state vector as shown by 

\begin{equation}
    \textbf{z} = \textbf{Hx} + \textbf{e}.
\end{equation}

The state estimation problem is to find the best fit estimate of $\hat{\textbf{x}}$ corresponding to the measured power flow values of $\textbf{z}$. Under the most widely used estimation approach, the state variables are determined by minimisation of a WLS optimisation problem as 

\begin{equation} \label{chisquaredfull}
 \underset{x}{\mathrm{min}}\,J(\textbf{x}) =  (\textbf{z}-\textbf{Hx})^T\textbf{W}(\textbf{z}-\textbf{Hx}).
\end{equation}
\textbf{W} is a diagonal $m \times m$ matrix consisting of the measurement weights. These weights can represent meter accuracy, reliability or simply engineering judgment about the relative importance of that particular measurement. The solution for a minimal $\textbf{J(x)}$ can be analytically obtained by taking the 1st derivative with respect to $\textbf{x}$ and solving for 0, yielding $\hat{\textbf{x}}$ defined by

\begin{equation} \label{estimate full}
    \hat{\textbf{x}} = (\textbf{H}^T\textbf{W}\textbf{H})^{-1}\textbf{H}^T\textbf{W}\textbf{z}.
\end{equation}

\subsection{Bad Data Detection}
The current approach in power systems operation for bad data detection is to use the 2-norm of the measurement residual. The residual $\textbf{r}$ is defined by the difference between the measured power flow values of $\textbf{z}$ and the value calculated from the estimated state values $\hat{\textbf{x}}$ and the known topology matrix $\textbf{H}$ 

\begin{equation} \label{residual}
    r =  ||\textbf{z} -\textbf{H}\hat{\textbf{x}}||_2.
\end{equation}

Assuming the state variable $\textbf{x}$ errors are random, independent and follow a normal distribution with mean zero and unit $\mathcal{N}(0,\sigma^2)$, a chi-squared distribution model $\chi^{2}_{m-n,\alpha}$ with $m-n$ degrees of freedom and $\alpha$ the confidence interval (typically 0.95 or 0.99) can be applied to define the detection threshold as
\begin{equation}
    \eta = \sigma \sqrt{\chi^{2}_{m-n,\alpha}}.      
\end{equation}

If $r_t>\eta$ BDD alarms will trigger and the system operator will discard the result, removing the elements from the residual calculation with large values.

\subsection{Attack Vectors}
In the case of an infinitely resourced and knowledgeable attack, the attacker can gain full access to the metering infrastructure of the power system and change measured power flows in almost any desired manner. However, the attacker will still wish to remain undetected by bypassing BDD. Considering an attack vector  $\textbf{a} \in \mathbb{R}^{m\times1}$ representing the change added to the measurements, the measurement vector under attack $\textbf{z}_a$ is

\begin{equation}
    \textbf{z}_a = \textbf{z} + \textbf{a}.
    \label{zattack}
\end{equation}

 As demonstrated in previous research, with a known system topology matrix $\textbf{H}$, it is trivial to create a stealthy attack vector. The attacker can choose any linear combination of $\textbf{H}\textbf{c}$ where $\textbf{c} \in \mathbb{R}^{n\times1}$. The vector $\textbf{c}$ can be selected so as to have the desired impact on the state vector $\textbf{x}$. With the attack vector $\textbf{a}$ shown by

\begin{equation}
    \textbf{a} =  \textbf{H}\textbf{c}.
\end{equation}

In most power systems $\textbf{H}$ is sparse. This means that most individual $c$ changes will correspond to only a few meter measurements. For the AC system this can also be generalised using partial derivative matrix $\textbf{J}$. In practice, the relative risk of these two models will be very similar and largely interconnection dependent.

\subsection{Moving Target Defence}
As shown previously, these types of attacks can be exposed using topology based defences known as MTD. We show in \cite{Higgins2020StealthySystems} that given an attack vector $\textbf{a}=\textbf{Hc}$ we can express the new residual $r_n$ in terms of the change in topology, size of the attack vector, WLS minimisation and power flow profile such that

\begin{equation} \label{residualhat1}
          r_n = \|(1 - \textbf{H}_n\textbf{F}_n)\textbf{z}+(1-\textbf{H}_n\textbf{F}_n)\Delta\textbf{H}\textbf{c}\|_2.
\end{equation}

Where $\textbf{F}_n$ is the WLS minimisation factor and $\textbf{H}_n$ is the new, post MTD topology. There is an assumption that attack vectors will be based on the original topology $\textbf{H}$. Generally, a system operator will aim to set this new residual value above his current alarm limits to ensure detection of FDI attacks. However, in practice, the attack vector cannot be known beforehand and MTD implementation can be costly and so understanding the exact level of MTD to apply can be difficult. 

\section{Cyber-Physical Threat Model}

\subsection{Attacker Assumptions}

We make some assumptions about the prospective attacker which help define our risk assessment model. We outline here the assumptions behind both the intrusion and system change elements of the attack.   

\begin{itemize}
    \item \textbf{System Intrusion} - the attacker is attempting an FDI style attack and will capture meter measurements in the required sub-graph. His intrusion cost will be the cost of compromising the meter set required to change a given bus measurement and they will seek to minimise this cost. The attacker can choose either to compromise the meter set, RTU or some combination to replicate the underlying attacking subgraph.  
    \item \textbf{System Change} - once intruded, the attacker is attempting to create a simulated overload attack via false data injection to the power flow profile. The attacker will attempt to optimise his attack vector to this effect, using the smallest possible attack vector needed to overload the given line. 
    \item \textbf{Statistical Peak} - the attacker is conscious of the additional advantage peak loading can grant and will wait for such instant before initiating the attack. We reflect this by perform simulations on a 3 standard deviation statistical peak reflecting an attackers advantage gained in waiting for an overloaded moment.
\end{itemize}

Given these attacker aims, we first consider the intrusion risk in terms of (weighted) sub-graph capture cost. We then consider the ability of the SO to defend the system with MTD under the 'peak load' style conditions.

\subsection{Min Cost Point Capture Strategy}

Previously, most works have envisioned an infinitely resourced attacker. In practice, attackers will be constrained in what elements they can compromise. They will likely choose targets based on ease to compromise and will prioritise low-cost targets. In this article, we outline formulations for weighted and unweighted bus capture strategies. The unweighted number of meters to compromise $MuC_n$ where $k_m$ denotes whether a meter is present at the busbar or branch. This is represented by the number of non-zero terms in the column vector of $\textbf{H}$ for a given node $n$ and represents the number of meters (needed to compromise) in order to attack stealthily. This represents a simply unweighted cost which is shown by

\[
    k_m= 
\begin{cases}
     1,& \text{if }\text{col}_{n}(H_{n,m})> 0\\
    0,              & \text{else}
\end{cases}
\]

\begin{equation}
    MuC_n =  \sum_{1}^m(k_m) .
\end{equation}

We can also add considerations of the difficulty to capture a given edge by adding a graph weighting. It makes sense given the co-existence of new and legacy measurement equipment in the power system. This can be represented with weightings for edges. For each edge $E$ of graph $G$ there is an associated weight $w(m)$. This can represent either line redundancy or more meters, which are more resilient to attack. The weighted cost of meters $MwC$ is shown by

\[
    p_m= 
\begin{cases}
     w(m),& \text{if } \text{col}_{n}(H_{n,m})> 0\\
    0,              & \text{else}
\end{cases}
\]

\begin{equation}
    MwC_n =  \sum_{1}^m(p_m) .
\end{equation}

Unlike in previous models, our proposed intrusion model the attacker can compromise either the metering components (similar to previous interpretations of FDI intrusion), the RTUs or some combination in order to replicate the needed subgraph. We outline these approaches in the next subsection.

 \begin{figure}[h]
 \centering
 \includegraphics[width=3.5in]{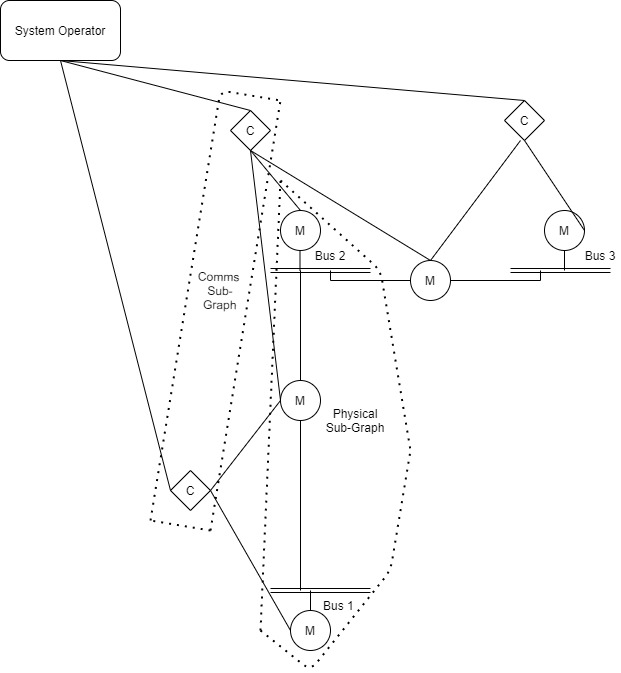}
 \caption{Cyber-Physical network for 3-bus system with alternative cyber and physical capture attack strategies.}
 \label{commsandphysattack}
 \end{figure}

 \begin{figure*}[h]
 \centering
 \includegraphics[height=3.0in]{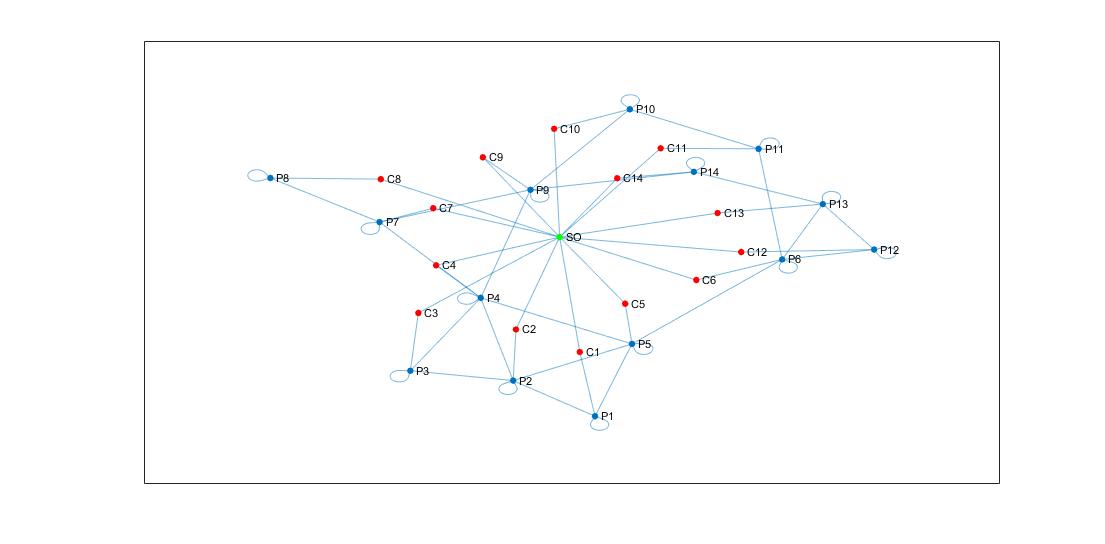}
 \caption{IEEE 14-BUS Cyber-Physical graph representation.} 
 \label{14buscps}
 \end{figure*}

 The weighted cost functionality also allows us to incorporate effects like the DoS style attacks as seen previously in \cite{Pan2019CyberEstimation}. 
 
 \subsection{State Capture Strategies}
 
\subsubsection{Meter Attack Criteria}
For a given node/state wishing to be attacked under the physical system the following is required to remain hidden: 
\begin{itemize}
    \item The self edge for the target node is captured and measurements changeable
    \item All edges emerging from the target node are captured
    \item The neighbour nodes of the target node self edges are also captured 
\end{itemize}

\subsubsection{RTU Attack Criteria}
For a given node/state wishing to be attacked under the communications network the following is required to remain hidden: 
\begin{itemize}
    \item The attacker would need to capture the RTUs (equivalent to capturing the network bus and adjacent power flow meters) associated with the physical attack. 
    \item The same physical sub-graph comprising of all the local measurements needs to be satisfied but through the capture of the upstream RTU which (usually) holds multiple meter measurements. 
\end{itemize}

\subsubsection{Combined Attack Criteria}
For a given node/state wishing to be attacked under the communications network the following is required to remain hidden: 
\begin{itemize}
    \item The attacker would need to capture some combination of RTUs and individual meters in order to satisfy the original "meter attack" criterion.
\end{itemize}

These attack options are shown in Figure \ref{commsandphysattack} for a 3 bus system. The physical attack sub graph requires capturing a number of system level branches while the communications strategy allows capturing of just 2 upstream nodes. Alternatively, the attack can opt for the combination attack, capturing one of the upstream RTU nodes and the left-over meter. 14-bus system representation is shown in Figure \ref{14buscps}.

\subsection{Physical Attack Risk}
In the past, the cost of attacking a particular busbar via FDI has been assessed in terms of the cost intrusion i.e. which meters are needed to manipulate a certain state stealthily. However, this method of assessment fails to consider MTD and the current system state. To assess the ability to attack a system a model should also consider the impact active detection will have on the system residual in the presence of an attack. As we've shown previously, in \cite{Higgins2020EnhancedDefence} meter value deviation from expected values can increase the chance of attack detection via anomaly detectors. We consider that the required level of MTD to protect a system is an important consideration and this level of MTD will be dependent on the size of the attack vector $\textbf{a}$. Meanwhile, we must also consider that (usually) $\textbf{c}$  will not be known ahead of the attack so a true max-min optimisation based on the attack vector will not be possible. Therefore, from the defender perspective, it is better to base risk calculations on known quantities. We consider that in order to perform a branch overloading attack the power flow profile of the system will have to be adjusted so that the power flow $\textbf{z}$ exceeds the capacity overhead $co$ . Therefore, the higher the $\textbf{co}$, the more tolerance for FDI attack an attacker has with respect to line overloading. 

\begin{equation} \label{residualhat}
     \textbf{co} = \|
     \textbf{z}_{cap}\|-\|\textbf{z}\|.
\end{equation}

Where $S$ is an $m \times m$ diagonal containing the power flow sign of $\textbf{z}$. Using this, we can get a set of $\textbf{c}$ values with the required branch change for overload. Setting all but the target branch in $\textbf{co}$ to 0, we can use this capacity overhead to get the set of voltage angles required to overload that branch such that

\begin{equation} \label{residualhat}
     \textbf{z}_{ol} = \textbf{z}+\textbf{H}\textbf{c} .
\end{equation}

We constrain each case so that only a single bus is being attacked (consistent with the minimal possible meter selection problem) and thereby setting $\textbf{c}^n=0$ for all except the current target bus. The attack vector is then given by

\begin{equation} \label{residualhat}
     \textbf{a}_{ol}^{n,m} = \textbf{H}\textbf{c} .
\end{equation}

Once the list of possible attack vectors have been evaluated for each bus, we use increasing magnitudes of available MTD in combination with a multi-variable optimisation of the topology $\textbf{H}$ to return the maximum residual for the level of installed MTD capacity.

\begin{equation}
\begin{array}{rrclcl}
\displaystyle \max_{\textbf{H}} & \multicolumn{3}{l}{ \| (\textbf{z}+\textbf{a}_{ol}^{n,m} ) -\textbf{H}(\textbf{H}^T\textbf{W}\textbf{H})^{-1}\textbf{H}^T\textbf{W}(\textbf{z}+\textbf{a}_{ol}^{n,m} ) \|_2}\\
\textrm{s.t.} & \textbf{H}<\textbf{H}_{limit}\\
\end{array}
\end{equation}

In practice, this optimisation can be performed quickly with limited processing power. This is because the optimisation is highly constrained. For one, the optimisation of topology only occurs over the relatively small range of the D-FACTs system limits. In addition, only a few branches will contribute to the residual calculation (namely those affected by the attacking subgraph). As power systems are sparse, this means that even in large systems only a small fraction of devices will need to optimised around for a given attacking subgraph. As a result of these factors, the overall boundary of optimisation in practice is very small and can be done quickly.    
The level of MTD divergence in absolute terms to evaluate each attack vector will be used as an assessment factor with regions requiring larger applications indicating an easier attack opportunity. We can use the WLS multiplier to find the relative level of divergence $DIV$ such that

\begin{equation} \label{residualhat}
     DIV=\| \textbf{H}(\textbf{H}^T\textbf{H})^{-1}\textbf{H}^T\textbf{H}_{mtd} \| .
\end{equation}

We use this required level of divergence to denote regions of higher risk with respect to FDI. Areas with high levels of $\textbf{CO}$ can be defended with very low levels of MTD applied, while regions operating at capacity can be overloaded with almost no FDI change and are therefore more difficult to protect using MTD.

\section{Cyber-Physical Assessment Algorithm}

\subsection{Weighted Min Cost Communications}

We use various tools from the MATLAB grTheory package \cite{Ahmad2015DevelopmentSystems} to establish the respective communications, physical and possible combination subgraphs and then evaluate the respective weighting for all of them. Initially, the algorithm takes the power network and communications graphs as inputs. It identifies the underlying attacking subgraphs. It then uses the MATLAB "NCHOOSEK" function to outline the different possible combinations of RTUs or meters that can achieve this subgraph, subsequently weighing each possible combination.

 \begin{figure*}[h]
 \centering
 \includegraphics[width=7in]{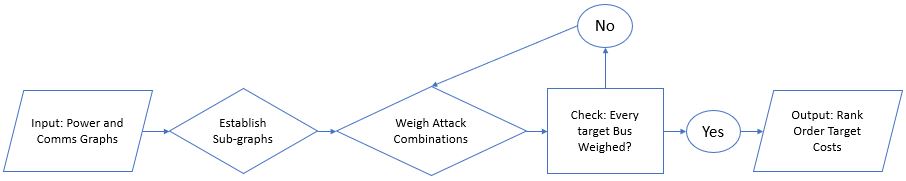}
 \caption{Outline of algorithm process for assessment of the weighted min meter target cost.}
 \label{cyberassessmentalgo}
 \end{figure*}
The cyber assessment algorithm runs as follows:
\begin{enumerate}
    \item To start, it takes the weighted graph inputs of the communications and physical meter networks.
    \item It then calculates the underlying physical subgraph (sgp) for each target bus.
    \item For the subgraphs, each possible RTU/meter combination to satisfy the attack is calculated.
    \item This is then repeated for each target bus and a rank order of target costs is established.
\end{enumerate}

\begin{figure}[h]
 \centering
 \includegraphics[width=3.5in]{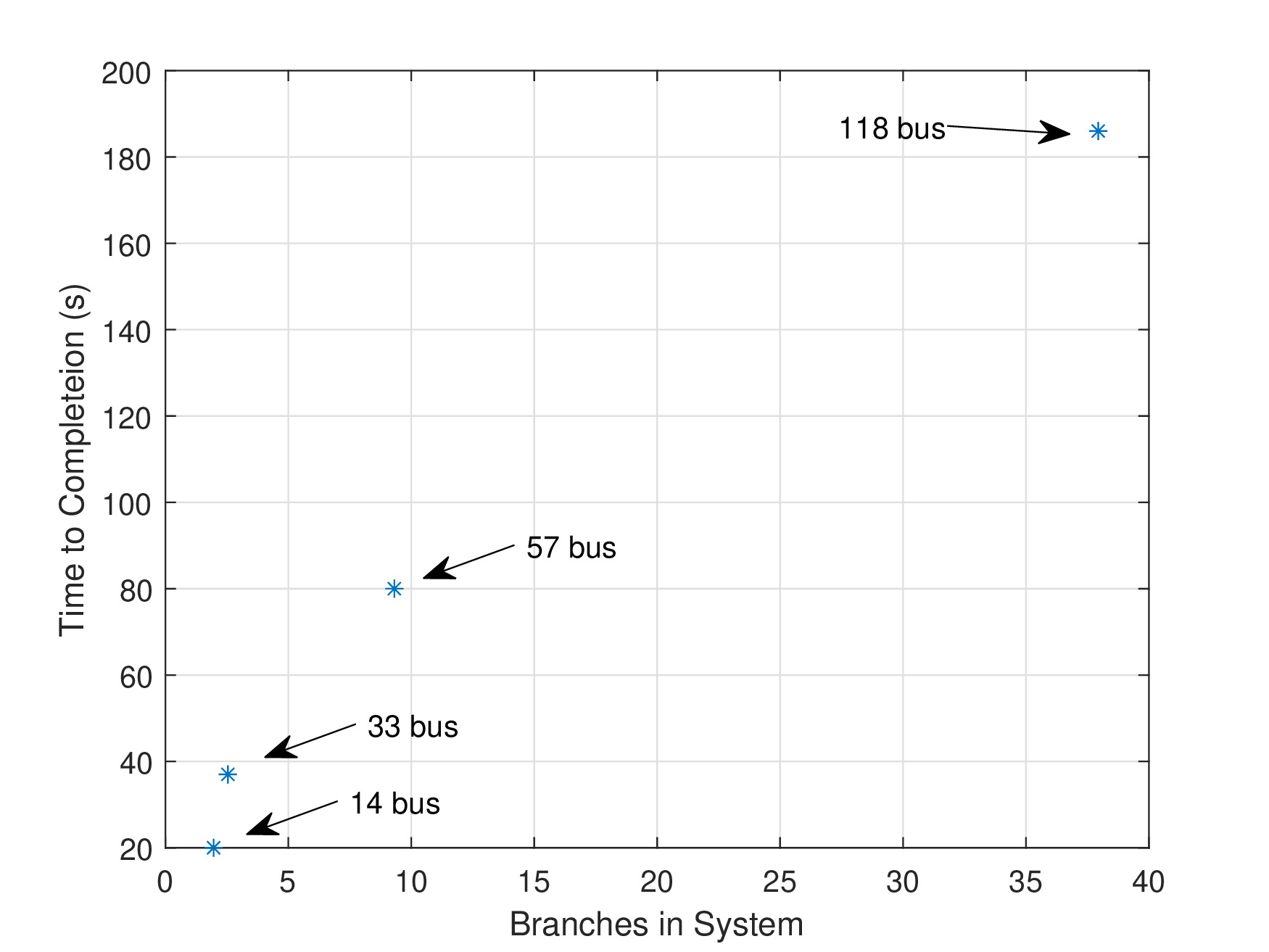}
 \caption{Time to completion of cyber-assessment algorithm process for systems of differing sizes.}
 \label{cyberassessmentalgotimes}
 \end{figure}

This algorithmic flow of this process is illustrated in Figure \ref{cyberassessmentalgo}. This algorithm can be applied quickly and simply. The time for varying system sizes is shown in Figure \ref{cyberassessmentalgotimes} with (as expected) linear time scaling. As power systems are sparse, even large systems would have broadly linear scaling. The exception to this would be systems where the level of interconnection grows with the system size, such as in the case of "complete" graphs.

\subsection{MTD based Physical Vulnerability Algorithm}
Cyber assessment can effectively model the risk of intrusion into a power system with respect to FDI attack vulnerabilities. However, in order to consider the post-intrusion consequences we need to consider the underlying power system model. In this post-intrusion assessment, we analyze the risk of a branch overload via FDI attack. One way to defend against FDI post-intrusion is using MTD. However, MTD is costly to operate and SOs seek to minimise their overall application of MTD. Large FDI attack vectors require less MTD to evaluate and are easier to identify. It follows that regions which require more MTD to protect have a higher inherent risk to FDI as they are harder to protect post-intrusion. Here we outline a method of assessing this risk. To do this we use the MATLAB multi-variable optimisation package FMINCON to establish the maximum residual value for a given level of MTD capacity (ranging from 1-50\% of base branch inductance). Initially, the algorithm takes the topology, MTD limits, power flows and power limits as inputs. Based on the power flow limits, minimum potential attack vectors are constructed from voltage angle adjustments for each branch. We then use the optimal MTD algorithm, with increasing capacity, to identify those regions which require the most overall MTD (as a \% of the base) to protect.     

The physical vulnerability algorithm runs as follows:

\begin{enumerate}
    \item The underlying power system data is taken as inputs namely; capacity, power flows, network topology and MTD limits.
    \item For each branch the overloading minimum change in voltage angles are calculated. 
    \item For each given attack vector and MTD capacity the maximum possible residual is calculated.
    \item The rank order of bus targets is established using the MTD divergence figure.
\end{enumerate}

\begin{figure*}[h]
 \centering
 \includegraphics[width=6in]{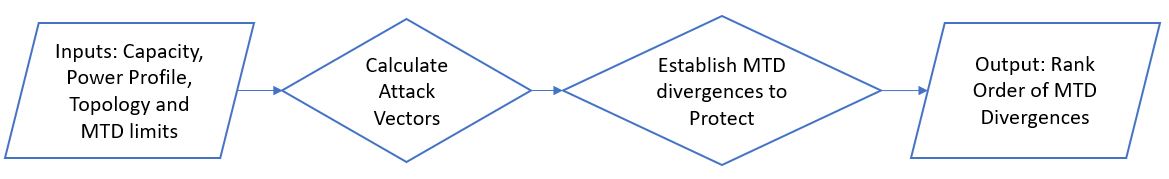}
 \caption{Outline of algorithm process for physical risk assessment using MTD divergence and line capacities.} 
 \label{physicalassessmentalgo}
 \end{figure*}

This algorithmic flow of this process is illustrated in Figure \ref{physicalassessmentalgo}.

\subsection{Statistical Load Peak}

We also consider the peak attack point for a respective attacker. Often, attackers can remain hidden for many months when intruding a system. Therefore it makes sense to operate under the assumption that the attacker will wait for the opportune moment to attack. We represent this opportune moment as a statistically significant load of 3 standard deviations from the mean such that

\begin{equation} \label{residualhat}
     \textbf{z}_{s} = \overline{\textbf{z}}+3\textbf{SD}_z.
\end{equation}

Where $\textbf{SD}$ is a vector of standard deviations of $z$ branch and bus values.

\begin{figure}[h]
 \centering
 \includegraphics[width=3.5in]{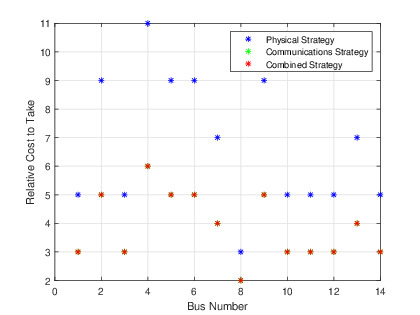}
 \caption{Weighted cost of each strategy with Node meters, Branch meters and RTUs equal to a weighted cost of 1} 
 \label{cybercase14normal}
 \end{figure}

  \begin{figure}[h]
 \centering
 \includegraphics[width=3.5in]{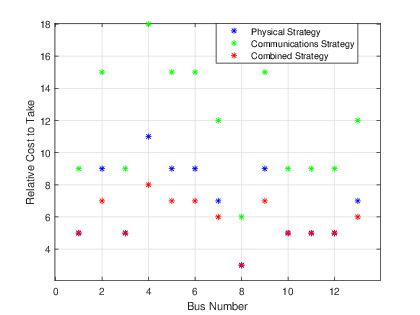}
 \caption{Weighted cost of each strategy with Node meters, Branch meters equal to 1 and RTUs equal to a weighted cost of 3} 
 \label{cybercase14normalthreetimescomms}
 \end{figure}


\section{Results \& Analysis}

This section shows the results of the proposed risk assessment strategies on both the standard IEEE 14-Bus test system \cite{ChristiePowerArchive}. All grid simulations were implemented using the MATPOWER toolbox in MATLAB \cite{Zimmerman2011MATPOWER:Education} and performed using Intel Core i7-7820X CPU with 64GB of ram running on a Windows 10 system.
\subsection{IEEE-14 Bus System Cyber}
In Figure \ref{cybercase14normal} we show the min target costs under the assumption of a flat cost of 1 for both meters or RTUs. From this graph it is clear that the communications-only strategy is always the most efficient under the assumption that the devices are of equal difficulty in capture. The reason that this effect occurs is because the RTUs sit upstream of meter measurements and thus have control of downstream meter measurements i.e. each RTU effectively has equivalence in capture to multiple meter measurements. Therefore, an attacker can replicate attacking subgraphs by attacking fewer of these upstream nodes rather than the meters directly. This is because this first example is an unweighted, non-reinforced model. In reality, most SOs would be aware that the RTUs represent a better target to the attacker due to this relationship (even prior to risk modelling) so we accept that this specific scenario is unlikely in a practical system. RTUs will probably have embedded defences against intrusions in place that will likely mean the RTUs will have higher levels of intrusion protection than base meter measurements. We illustrate this in Figure \ref{cybercase14normalthreetimescomms}. In this graph, we introduce a weighted cost for RTU capture of factor 3 times the direct meter compromise. In practice, weighted reinforcement at the RTU level is a more realistic assumption and makes sense from a system operation perspective. RTUs control multiple functions and have downstream capabilities. The value in their protection will be more crucial than simple meter measurements, which only provide telemetered measurements. Crucially, the weighting of RTUs in this manner shows the emergence of combined strategies emerging as the most efficient use of resourcing. This makes sense as these combined approaches allow an attacker to utilise well-connected RTUs and isolated meters to complete underlying attack subgraphs.

\subsection{IEEE-14 Bus System Physical}

\begin{figure*}[!h]
 \centering
 \includegraphics[width=7.5in]{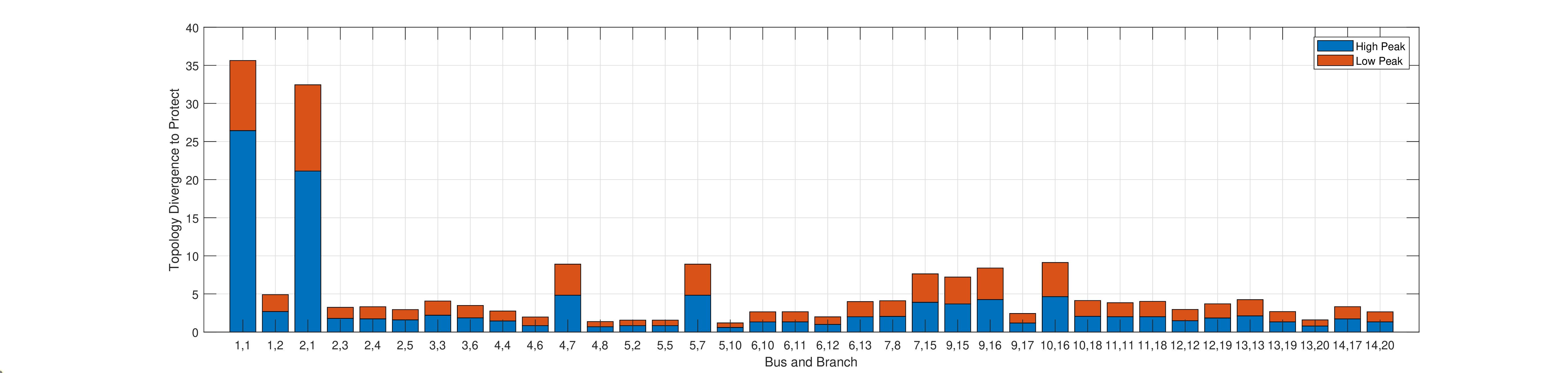}
 \caption{Absolute topology divergence to evaluate an attack for each bus and corresponding branch overload under a statistical peak of 3 standard deviations operating conditions high and the same boundary lower.} 
 \label{div3sdstatpeak}
 \end{figure*}

\begin{figure*}[!h]
 \centering
 \includegraphics[width=7.5in]{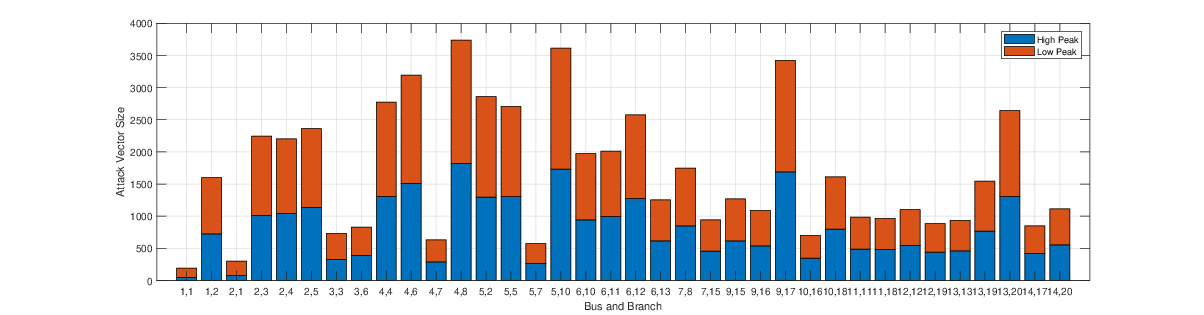}
 \caption{Absolute required relative attack vector size to overload a line an attack for each bus and corresponding branch overload under a statistical peak of 3 standard deviations operating conditions high and the same boundary lower.} 
 \label{asize}
 \end{figure*}

In Figure \ref{div3sdstatpeak} we explore the impact of increasing MTD divergence on system residual and detection. As discussed previously, we use the optimised min attack vector required to overload a line within each attacking subgraph. We note that in branches where the natural power flow profile is close to the network capacity, only smaller attacking vectors are needed in order to achieve the simulated overload. This makes sense from a system operation perspective as areas close to limits require only small changes to overload a branch in excess of capacity limits. We should expect to see that branches with higher peak flows have higher MTD divergence requirements (as the attack vectors are smaller and harder to evaluate with MTD). Indeed, we see that for regions with high overload capacity (low branch power flow relative to capacity) only a via minor application of MTD based divergence is required to evaluate the optimal branch overloading attack. However, as we can see in Figure \ref{asize}, there is an inverse relationship between the size of attacking vector and level of MTD required to protect against the attack. This is particularly clear when the high vs low peak results are observed. For example, bus 1-2 operates at the closest point to the branch capacity and we see large levels of MTD divergence are required to protect this bus sufficiently from cyber attacks.  Significantly higher levels of overall divergence are required to defend the system which means these points are comparatively susceptible to FDI based changes. As the average size of attack vector to breach the system is lower, the level of divergence required to evaluate a FDI attack increases. This makes buses with these close to overloaded branches relatively better targets than other regions where the attack vector has to be large (and hence more easily evaluated).

 \subsection{IEEE-14 Bus System Cyber- Physical}
We now consider attack targets in terms of their whole system risk. Bus 8 is by far the most vulnerable target in every non-reinforced model. The relatively low interconnected means it has the lowest capture cost of all the available buses with just 2 components needed in order to compromise and gain a stealthy intrusion. Also, the lack of interconnection also means that MTD protocols are ineffective as MTD requires at least 2 interconnections within an attacking subgraph to drive changes to the residual under attack. This means that despite the low line power flows, MTD is ineffective.  In light of this consideration, bus 8 should be the priority busbar for intrusion based reinforcement and physical reinforcement should be ignores which provides no benefits to this busbar. This particular case has crucial implications for systems such as the IEEE 33-bus distribution style network due to their lower levels of interconnectivity resulting from the tree style topology. The vast majority of attack points in these types of systems would gain no defensive advantage with MTD and so intrusion based defences should be prioritized. Similarly, in the case of busbar points 1 \& 2 there are large defensive MTD requirements in order to protect the branch 1-2 measurement. While this branch is defensible with MTD from an absolute cost perspective it is likely better to consider enhancing intrusion protection. Predictably, highly connected buses have some innate protection when it comes to FDI  style attacks. Busbar 6 for example has 4 interconnections, this means an innately higher system protection from an intrusion perspective with 9 underlying meters needing to be captured to commit stealthy changes.

\section{Conclusions \& Further Work}
In this work, we developed a risk assessment framework for false data injection attacks. Our assessment criteria considers weighted graph assessment of the cyber-vulnerabilities in combination with a residual based assessment of the physical system with relation to MTD. This framework provides, for the first time, an intrusion and change introduction model for risk assessment. This model first considers the weighted minimum cost of intrusion into the network subgraph by both RTU, meter and combined means. Second the model considers residual under the minimum overloading attack in the presence of MTD to show how defensible the targets are. Simulations are performed under the assumption of the peak load system in order to replicate the attacker's ability to wait for opportune moments to strike. To date, most work in the field of FDI attacks has occurred purely in simulation. However, there is a need to take these attack types from the simulation realm to real life simulation on cyber-physical system testbeds such as the one outlined in \cite{Adepu2018EPIC:Systems}.

\bibliographystyle{IEEEtran.bst}

\bibliography{main.bbl}

\end{document}